\documentstyle[12pt,a4]{article}
\newlength{\bredde}
\def\slash#1{\settowidth{\bredde}{$#1$}\ifmmode\,\raisebox{.15ex}{/}
\hspace*{-\bredde} #1\else$\,\raisebox{.15ex}{/}\hspace*{-\bredde} #1$\fi}
\newcommand{\beq}{\begin{equation}}
\newcommand{\eeq}{\end{equation}}
\newcommand{\bea}{\begin{eqnarray}}
\newcommand{\eea}{\end{eqnarray}}

\def\gtwid{\raise.3ex\hbox{$>$\kern-.75em\lower1ex\hbox{$\sim$}}}
\def\ltwid{\raise.3ex\hbox{$<$\kern-.75em\lower1ex\hbox{$\sim$}}}
\def\AA{\widehat A}
\def\pp{\widehat \Phi}
\def\DD{\widehat D}
\def\FF{\widehat F}
\def\vv{\widehat v}
\def\tr{{\rm tr}}

\title{MASS GENERATION IN THREE DIMENSIONS}
\author{{\sc M. Oleszczuk
}\thanks{E-mail:
Oleszczuk@vax1.rz.uni-regensburg.d400.de} \\
Institut f\"ur Theoretische Physik \\ Universit\"at Regensburg\\ D-93040
Regensburg, Germany
}
\date{}
\begin{document}

\maketitle

\begin{abstract}
I consider the dilute monopole gas expansion of the three
dimensional Yang-Mills-Higgs system in the symmetry broken
phase. The functional determinants which occur in such an
expansion are computed in the heat kernel approximation for
an arbitrary $SU(N)$ gauge group. Explicit expressions for the
gauge boson mass in the unbroken gauge sector and the string
tension are obtained for the $SU(2)$ gauge model and are
evaluated numerically. The results show a strong dependence
on the ratio $m_{\rm Higgs}/m_W$.
\end{abstract}

\vspace{-220mm}
\vfill\hfill
\vbox{
\hfill TPR-94-35\null\par
\hfill hep-th/9412049}\null
\vspace{220mm}

\newpage
\baselineskip 20pt

\section{Introduction}

The dynamics of three and four dimensional systems differ
substantially. A characteristic difference is that the infrared
modes play a more important role in the lower dimensional case. On
the one hand this simplifies the perturbative investigations
since the ultra-violet regime is less important and, by that,
the separation of the non-physical, ultra-violet divergences
becomes easier. On the other hand, however, the infrared modes
may render the usual perturbation expansion divergent order by
order in the infrared. In such a case we are forced to perform
a partial resummation of the perturbation expansion. The
resulting improved perturbation expansion is based on certain
quasi-particles with screened interactions and may be very
different from the original perturbation expansion.

These phenomena can nicely be studied by means of the three
dimensional $SU(N)$ Yang-Mills-Higgs system in the symmetry
broken phase. The action of this model is given by
\begin{equation}
S[A^a_i,\Phi^a]=\int d^3x\left({1\over 4}F^a_{ij}F^a_{ij}
+{1\over 2}(D_i\Phi)^a(D_i\Phi)^a
+{\lambda\over 8}\left(\Phi^{a2} - v^{a2}\right)^2\right)\;,
\label{act}
\end{equation}
where
\begin{eqnarray}
F^a_{ij} &=& \partial_i A^a_j - \partial_j A^a_i
+ef^{abc}A^b_i A^c_j\;,
\nonumber \\
\nonumber \\
(D_i \Phi)^a &=& \partial_i \Phi^a + ef^{abc}A^b_i \Phi^c\;,
\label{Fe}
\end{eqnarray}
$f^{abc}$ are the structure constants of the $SU(N)$ gauge group and
the coupling constants $e^2$ and $\lambda$ have the dimension of an
energy. Since the theory is superrenormalizable one only needs a single
counterterm for the mass renormalization of the Higgs field $\Phi$.
But the perturbation expansion is order by order divergent in the
infrared and thus renders the theory essentially nonperturbative.

Fortunately, the action (\ref{act}) possesses stable localized extrema,
the static Yang-Mills-Higgs  monopole configurations of the
corresponding four dimensional gauge model \cite{mona,monb,monc,mond,mone}.
The polarization of a dilute gas of such pseudo-particles screens the
long range magnetic Coulomb forces so that the semiclassical expansion
yields an infrared stable approximation \cite{poly}. According to this
expansion the electric charges are confined by a linear potential and the
gauge bosons acquire a non-vanishing mass.

Originally, the reason for studying the $SU(2)$ dilute monopole gas
expansion was the fact that QED in the Georgi-Glashow model is a simple
example of a compact $U(1)$ gauge theory. Polyakov's work
\cite{poly} in 1976 had initiated an extensive study of the three
dimensional compact $U(1)$ gauge theory and in particular of a lattice
version of this $U(1)$ theory that is obtained by using the Villain
approximation of the Wilson action \cite{lattice}. The primary goal of
these lattice studies was and still is to reveal the phase structure
of such theories by studying the behavior of quantities such as the
string tension as a function of the inverse temperature $\beta=1/e^2a$
where $a$ is the lattice spacing.

But there are various other important aspects of the three dimensional
model (\ref{act}) which require a detailed investigation of the dilute
monopole gas expansion in the context of the model itself. One certainly
is the fact that it resembles the effective model for the high temperature
phase of QCD \cite{olpa, olpb, olpc}. Through dimensional reduction
of the four dimensional gauge theory of strong interactions at finite
temperature one arrives at a three dimensional gauge model in which the
zero-component of the gauge field plays the role of the Higgs field.
The Higgs sector is screened by the Debye mass whereas the gauge sector
is infrared divergent. Therefore conventional perturbation expansion is
inapplicable. To overcome this difficulty one has to perform a
certain partial resummation of infinitely many Feynman diagrams. Such a
resummation could be very similar to the dilute monopole gas expansion.

Another important issue is the question of the true vacuum. In most
computations of the effective potential only the trivial constant
background field is considered. However it is well-known since the work
of Coleman \cite{cola, colb} that nontrivial saddle point configurations of
the underlying theory may alter these results fundamentally in the sense
that what appears to be the vacuum state is actually unstable. For
instance if we consider the three dimensional Yang-Mills-Higgs model
in the symmetrical phase the one loop effective potential obtained with a
constant background field shows no symmetry breaking effects. This may
change if we also include non-trivial saddle point configurations in
the computation. Such nontrivial saddle point configurations arise because
of the explicit symmetry breaking source term that we commonly introduce
in order to study spontaneous symmetry breaking. The form of these
configurations is very similar to the 't Hooft-Polyakov monopoles and
their contributions to the effective potential can be computed in the
dilute gas expansion. However, in order to investigate such a possibility
of spontaneous symmetry breaking we first must have a complete description
of the dilute monopole gas.

The present paper contains further details of the dilute monopole gas
expansion which was sketched in Ref. \cite{poly}. Its goal is to provide
the formulae for the quantitative analysis of the expansion. This
is achieved by an approximate computation of the 1-loop determinants
which occur in the monopole partition function. We regularize the
determinants with the help of the $\zeta$-function regularization and
then evaluate the resulting expressions in the heat kernel approximation
\cite{taubes, dyak}. This enables us to present explicit expressions for
the gauge boson mass in the unbroken $U(1)$ gauge sector and the string
tension in the $SU(2)$ gauge model.

The numerical evaluation of the formulas shows that the $U(1)$ gauge
boson mass and the string tension strongly depend on the ratio
$\lambda/e^2=m^2_{\rm Higgs}/m^2_W$ where $m_W$ is the heavy vector
boson mass. In both limits, $\lambda/e^2\to0$ and $\lambda/e^2\to\infty$,
the $U(1)$ gauge boson mass and the string tension vanish but physics
is different. We shall argue that in the Prasad-Sommerfield limit
$\lambda/e^2\to0$ the symmetry is restored whereas the limit
$\lambda/e^2\to\infty$ corresponds to a non-confining Higgs phase
with a massless $U(1)$ gauge boson.

Since there exist several misprints in the original literature
and since the precise expressions which arise in the formulation of
the dilute monopole gas expansion are important for our considerations
we find it appropriate to present the dilute monopole gas expansion
in greater detail. In Section 2 we rederive the one-monopole partition
function for the $SU(N)$ Yang-Mills-Higgs theory. The evaluation of
the $SU(N)$ one-loop determinants with the help of the
heat kernel method is subject of Section 3. In Section 4 we give the
precise expressions for the $SU(2)$ dilute monopole gas partition function,
the gauge boson mass in the unbroken gauge sector and the string tension.
Numerical results for the $SU(2)$ gauge model are presented in Section 5
and a summary is given in Section 6.

\vspace{1.0cm}

\section{The $SU(N)$ one-monopole partition function}

The general presupposition for the semi-classical expansion is the
existence of non-trivial saddle point configurations of the
underlying classical action. Since the action (\ref{act}) of the three
dimensional $SU(N)$ Yang-Mills-Higgs system is just the energy of
the static configurations of the corresponding four dimensional
system, the static solitons of the four dimensional theory -- the
Yang-Mills-Higgs magnetic monopoles -- are the saddle points or
pseudo-particles of the three dimensional model.

In order to obtain the one-monopole contribution to the functional
integral,
\begin{equation}
Z=\int {\cal D}[A^a_i]{\cal D}[\Phi^a]
\exp\left(-S[A^a_i,\Phi^a]\right)\; ,
\label{Z}
\end{equation}
in the one-loop approximation one writes the fields $A^a_i$ ($\Phi^a$) as
a sum of the classical one-monopole field $\AA^a_i$ ($\pp^a$) and a
quantum fluctuation $a^a_i$ ($\phi^a$) and expands the action up to
terms quadratic in the quantum fluctuations. As usual the functional
integration over the quantum fluctuations requires some gauge-fixing. A
suitable choice is the background-field gauge defined by the
gauge-fixing action
\begin{equation}
S_{\rm gf}[A^a_i,\Phi^a] = {1\over2}\int d^3x
\left(\DD_i^{ab}\left(A^b_i - \AA^b_i\right)
+ ef^{abc}\pp^b\left(\Phi^c-\pp^c\right)\right)^2\;.
\label{sg}
\end{equation}
The corresponding Faddeev-Popov determinant is given by
\begin{equation}
\det\left[{\cal M}_{\rm FP}\right]
= \det\left[\DD^{ac}_i D^{cb}_i
+ e^2f^{acd}f^{dc'b}\pp^c\Phi^{c'}\right]\;,
\label{dg}
\end{equation}
which, in a one-loop computation can be approximated by
\begin{equation}
\det\left[{\cal M}_{\rm FP}\right]
\approx\det\left[\DD^{ac}_i\DD^{cb}_i
+ e^2f^{acd}f^{dc'b}\pp^c\pp^{c'}\right]\;.
\label{dgg}
\end{equation}

If one performs the standard algebraic manipulations and employs
the classical field equations
\begin{eqnarray}
(\DD_i\FF_i)^a &=& ef^{abc}\pp^b(\DD_j\pp)^c\;,
\nonumber \\
\nonumber \\
(\DD_i\DD_i\pp)^a
&=& {\lambda\over2}\left(\pp^{a2}-v^{a2}\right)\pp^a\;,
\label{fe}
\end{eqnarray}
as well as the Bianchi identity for the $SU(N)$ structure constants
one arrives at the following expression for the functional integral
in the one-monopole sector
\begin{equation}
Z_1=\int {\cal D}[a^a_i]{\cal D}[\phi^a]
\det\left[{\cal M}_{\rm FP}\right]
\exp\left(-S_{\rm m}[\AA^a_i,\pp^a]-S_{\rm quadr}[a^a_i,\phi^a]\right)\;.
\label{ZZ}
\end{equation}
Here, $S_{\rm m}[\AA^a_i,\pp^a]$ is the classical one-monopole action,
\begin{equation}
S_{\rm m}[\AA^a_i,\pp^a]=\int d^3x\left(
 {1\over 4}\FF^a_{ij}\FF^a_{ij}
+{1\over 2}(\DD_i\pp)^a(\DD_i \pp)^a
+{\lambda \over 8}\left(\pp^{a2} - v^{a2}\right)^2\right)\;,
\label{sc}
\end{equation}
and $S_{\rm quadr}[a^a_i,\phi^a]$ that part of the full action which is
quadratic in the quantum fluctuations:
\begin{equation}
S_{\rm quadr}[a^a_i,\phi^a] = {1\over2}\int d^3x
\left(\matrix{a^a_i,&\phi^a}\right)
\left(\matrix{{\cal M}^{ab}_{ij}&-{\cal M}^{ab}_{i}\cr
              {\cal M}^{ab}_{j} & {\cal M}^{ab}\cr}\right)
\left(\matrix{a^b_j\cr \phi^b \cr}\right)\;,
\label{ssm}
\end{equation}
where
\begin{eqnarray}
{\cal M}^{ab}_{ij}
&=& -\DD^{ac}_k\DD^{cb}_k\delta_{ij}
- 2ef^{acb}\FF^c_{ij}
- e^2f^{acd}f^{dc'b}\pp^c\pp^{c'}\delta_{ij}\; ,
\nonumber \\
\nonumber \\
{\cal M}^{ab}_{i}
&=& 2ef^{acb}(\DD_i\pp)^c\; ,
\nonumber \\
\nonumber \\
{\cal M}^{ab}_{j}
&=& 2ef^{acb}(\DD_j\pp)^c\; ,
\nonumber \\
\nonumber \\
{\cal M}^{ab}
&=& -\DD^{ac}_k\DD^{cb}_k
+ {\lambda\over2}\left(\pp^{c2}-v^{c2}\right)\delta^{ab}
+ {\lambda}\pp^a\pp^b
-e^2f^{acd}f^{dc'b}\pp^c\pp^{c'}\; .
\label{Mm}
\end{eqnarray}

Apart from the gauge zero modes which are eliminated by the gauge-fixing
there are additional zero modes since the classical one-monopole solution
violates translational symmetry. The corresponding eigenfunctions of the
quadratic form (\ref{Mm}) are given by \cite{poly}
\begin{eqnarray}
{a^{(l)a}_j\choose \phi^{(l)a}}
&=& {1\over \sqrt{{\cal N}}}{\FF^a_{lj}\choose(\DD_l\pp)^a}\; ,
\nonumber \\
\nonumber \\
\nonumber \\
{\cal N} &=& \int d^3x\left(\FF^a_{ij}\FF^a_{ij} +
(\DD_i\pp)^a(\DD_i\pp)^a\right)\; .
\label{ef}
\end{eqnarray}
To eliminate these zero modes in the functional integral (\ref{ZZ})
we again use the standard Faddeev-Popov method of insertion of unity. The
appropriate decomposition of unity is
\begin{eqnarray}
1 &=& \det_{kl}\left[{1\over\sqrt{{\cal N}}}
\int d^3x\left(
{\delta\over\delta R_k}a^a_i\FF^a_{li} +
{\delta\over\delta R_k}\phi^a(\DD_l\pp)^a\right)\right]
\nonumber \\ & &
\nonumber \\ & &
\nonumber \\ & &
\qquad\times\int d{\bf R}
\prod_{l=1}^3\delta\left({1\over \sqrt{{\cal N}}}
\int d^3x\left(a^a_i\FF^a_{li}+\phi^a(\DD_l\pp)^a\right)\right)
\label{unity}
\end{eqnarray}
where ${\bf R}$ is the center of mass coordinate of the monopole. In the
one-loop approximation the determinant in (\ref{unity}) can easily be
evaluated and gives ${\cal N}^{3/2}$. Thus if we insert (\ref{unity})
in the functional integral (\ref{ZZ}) we obtain
\begin{eqnarray}
Z_1&=&\int d{\bf R}\; {\cal N}^{3/2}
\det\left[M_{\rm FP}\right]\exp\left(-S_{\rm m}[\AA^a,\pp^a]\right)
\nonumber \\ &&
\quad\times\int {\cal D}[a^a_i]{\cal D}[\phi^a]
\prod_{l=1}^3\delta\left({1\over \sqrt{{\cal N}}}
\int d^3x\left(a^a_i\FF^a_{li}+\phi^a(\DD_l\pp)^a\right)\right)
\nonumber \\ &&
\quad\qquad\times\exp\left(-S_{\rm quadr}[a^a_i,\phi^a_i]\right)\;.
\label{ZZZ}
\end{eqnarray}

Since no further zero modes exist we can now do the functional
integral over the quantum fluctuations. If we use Pauli-Villars
regularization the expression for the one-monopole contribution to
the functional integral in the one-loop approximation is given by
\begin{equation}
Z_1=\int d{\bf R}\; M^3{\cal N}^{3/2}
\det\left[{\cal M}_{\rm FP}\right]
\left(\widetilde{\det}\left[{\cal M}\right]\right)^{-1/2}
\exp\left(-S_{\rm m}[\AA^a_i,\pp^a]\right)
\label{ZZZZs}
\end{equation}
where $M$ is the Pauli-Villars regulator mass and ${\cal M}$ the quadratic
form defined in (\ref{ssm}) and (\ref{Mm}). In case of the quantum
fluctuation determinant we write $\widetilde{\det}$ instead of $\det$ to
indicate that the determinant has to be calculated with respect to the
non-zero modes of ${\cal M}$ only. Furthermore it is implied that both
functional determinants are regularized by the Pauli-Villars method and
normalized by the corresponding 'free' determinants, $\det[{\cal M}^0_{FP}]$
and $\det[{\cal M}^0]$. ${\cal M}^0_{FP}$ and ${\cal M}^0$ are obtained
from ${\cal M}_{FP}$ and ${\cal M}$ by setting $\AA^a_i = 0$ and
$\pp^a = v^a$.

Finally, it is useful to introduce the dimensionless
fields $\Phi^a\to v\Phi^a$ and $A^a_i\to vA^a_i$ as well as the
coordinates $x_i\to x_i/(ev)$, $R_i\to R_i/(ev)$ and $M\to Mev$
In terms of these dimensionless quantities the functional integral
(\ref{ZZZZs})
reads
\begin{equation}
Z_1=\left({m_W\over e^2}\right)^{3/2}\int d{\bf R}\; M^3
{\cal N}^{3/2}
\left({\det\left[{\cal M}_{\rm FP}\right]\over
\det\left[{\cal M}^0_{\rm FP}\right]}\right)
\left({\widetilde{\det}\left[{\cal M}\right]\over
{\det}\left[{\cal M}^0\right]}\right)^{-1/2}
\exp\left(-S_{\rm m}[\AA^a_i,\pp^a]\right)
\label{ZZZZ}
\end{equation}
where $m_W=ve$ is heavy vector boson mass,
\begin{equation}
S[\AA,\pp]
={m_W\over e^2}\int d^3x\left({1\over 4}\FF^a_{ij}\FF^a_{ij}
+{1\over 2}(\DD_i \pp)^a(\DD_i \pp)^a
+{\lambda \over 8e^2}\left(\pp^{a2} - 1\right)^2\right)\;,
\label{ss}
\end{equation}
\begin{eqnarray}
\FF^a_{ij} &=& \partial_i \AA^a_j - \partial_j \AA^a_i
+f^{abc}\AA^b_i \AA^c_j\;, \phantom{aaaaaaaaaaaaaaaaaaaaaaaaaaaa}
\nonumber \\
\nonumber \\
(\DD_i \pp)^a &=& \partial_i \pp^a + f^{abc}\AA^b_i \pp^c\;,
\label{FFF}
\end{eqnarray}
\begin{equation}
\nonumber \\
{\cal M}_{\rm FP}^{ab}=
\DD^{ac}_i\DD^{cb}_i + f^{acd}f^{dc'b}\pp^c\pp^{c'}\;,
\phantom{aaaaaaaaaaaaaaaaaaaaaaaaa}
\label{mfp}
\end{equation}
and
\begin{eqnarray}
{\cal M}^{ab}_{ij}
&=& -\DD^{ac}_k\DD^{cb}_k\delta_{ij}
- 2f^{acb}\FF^c_{ij} - f^{acd}f^{dc'b}\pp^c\pp^{c'}\delta_{ij}\; ,
\nonumber \\
\nonumber \\
{\cal M}^{ab}_{i}
&=& 2f^{acb}(\DD_i\pp)^c\; ,
\nonumber \\
\nonumber \\
{\cal M}^{ab}_{j}
&=& 2f^{acb}(\DD_j\pp)^c\; ,
\nonumber \\
\nonumber \\
{\cal M}^{ab}
&=& -\DD^{ac}_k\DD^{cb}_k
+ {\lambda\over 2e^2}\left(\pp^{c2}-1\right)\delta^{ab}
+ {\lambda\over e^2}\pp^a\pp^b
-f^{acd}f^{dc'b}\pp^c\pp^{c'}\;.
\label{Mmm}
\end{eqnarray}

With the help of the one-monopole result (\ref{ZZZZ}) one can now construct
a formal expression for the partition function of a dilute gas of infinitely
many monopoles (see Section 4). However, since we are interested in an
explicit expression for this partition function which then can be
investigated numerically we are forced to work out the one-loop
determinants. What we can say already at this point by only looking at the
above formulas for the one-loop determinants is that the result will depend
on the ratio $\lambda/e^2$, only.

\vspace{1.0cm}
\section{Calculation of the $SU(N)$ determinants}

The method we are going to use to evaluate the determinants occurring in
the expression for the one-monopole partition function is explained in
detail in Refs. \cite{taubes} and \cite{dyak}. It is based on the
observation that the ratio of two determinants,
$\widetilde{\det}{\cal A}$ and
$\widetilde{\det}{\cal B}$ with finite number of zero modes,
$n_{\cal A}$ and $n_{\cal B}$, can be written as
\begin{equation}
\ln\left({\widetilde{\det}{\cal A} \over
\widetilde{\det} {\cal B}}\right) =-\int_0^\infty {dt\over t}
\left[{\rm Tr}\left(e^{-t{\cal A}}-e^{-t{\cal B}}\right)
-n_{\cal A} + n_{\cal B}\right]\;.
\label{formel}
\end{equation}
In general this expression has to be regularized. The most suited
regularization scheme is the $\zeta$-function regularization
\cite{hawking} defined by
\begin{eqnarray}
\lefteqn{\ln\left({\widetilde{\det}{\cal A} \over
\widetilde{\det} {\cal B}}\right)_{\rm reg} =} &&
\nonumber \\ &&
\qquad=-\lim_{s\to0}\left\{{d\over ds}\left({M^{2s}\over\Gamma(s)}
\int_0^\infty dt t^{s-1}
\left[{\rm Tr}\left(e^{-t{\cal A}}-e^{-t{\cal B}}\right)
-n_{\cal A} + n_{\cal B}\right]\right)\right\}\;,
\nonumber \\ &&
\label{formelr}
\end{eqnarray}
where $M$is again the Pauli-Villars regulator mass.
The functional trace can easily be evaluated with respect to the
complete set of plane waves $\exp(-ip_j x_j)$. Then we only
have to shift all differential operators $\partial_j$ occurring in
${\cal A}$ and ${\cal B}$ by $ip_j$ and integrate over $d^dp$.
Consequently (\ref{formelr}) becomes
\begin{eqnarray}
\lefteqn{\ln
\left(
{ \widetilde{\det}{\cal A}\over\widetilde{\det}{\cal B} }
\right)_{\rm reg}
=-\lim_{s\to0}
\left\{
{d\over ds}
\left(
{M^{2s}\over\Gamma(s)}\int_0^\infty dt t^{s-1}
\right.
\right.} &&
\nonumber \\ &&
\nonumber \\ &&
\qquad\times
\left.\left.
\left[
\int d^dx \int {d^dp\over(2\pi)^d}
\tr
\left(
e^{-t{\cal A}(\partial_j\to\partial_j+ip_j)}-
e^{-t{\cal B}(\partial_j\to\partial_j+ip_j)}
\right)
-n_{\cal A}+n_{\cal B}
\right]
\right)
\right\}
\;,
\nonumber \\ &&
\label{ff1}
\end{eqnarray}
where we used "tr" instead of "Tr" to indicate that the trace in the
functional space has already been performed.

Mostly it is rather hopeless to compute expression (\ref{ff1})
exactly and one has to rely on a certain approximation. The so-called
"heat kernel" approximation is based on a power series expansion
in $t$ of the integrand occurring in (\ref{ff1}). For dimensional reason an
equivalent method is first to expand the exponential functions in powers of
the covariant derivative (or the classical background-field) and
then to integrate over the momentum. The approximation consists of
cutting off the power series expansion in $t$ at a suitable order
and replacing the upper limit of the $t$-integration by a finite
limit $t_0$ such that the integral is stationary at $t_0$.

In Ref. \cite{dyak} this method has been used to evaluate the determinants
occurring in the functional integral of the pure gauge theory in the
presence of an instanton background field. Thereby it turned out that
already an approximation in which only the first two nontrivial
terms of the expansion in $t$ are taken into account agrees
with the exact result within 3\%. A possible explanation of the
surprising accuracy of the approximation is that the first two nontrivial
terms are already the third and fourth order terms in the $t$-expansion.

Since the evaluation of the determinants becomes much more complicated
for the Yang-Mills-Higgs system than for the pure gauge theory we shall
restrict ourselves to the first two nontrivial terms in the
$t$-expansion, too. However, it should be mentioned that in our case
these contributions will result from the second and third order terms of
the $t$-expansion. Therefore we cannot expect a similar accuracy like in
the corresponding instanton computation of \cite{dyak} but we believe that
the approximation should still be reasonable.

Let us begin with the evaluation of the Faddeev-Popov determinant, i.e.,
\begin{equation}
\ln\left(
{\det[-{\cal M}_{\rm FP}]\over \det[-{\cal M}^0_{\rm FP}]}
\right)_{\rm reg} =
-\lim_{s\to0}\left\{{d\over ds}\left({M^{2s}\over\Gamma(s)}
\int_0^\infty dt t^{s-1}
{\rm Tr}\left(e^{t{\cal M}_{FP}}-e^{t{\cal M}^0_{\rm FP}}\right)
\right)\right\}\;.
\label{fadp}
\end{equation}
As described above we insert a complete set of plane waves and expand
the exponential $\exp(t{\cal M}_{\rm FP})$ in powers of the covariant
derivatives. This leads us to
\begin{eqnarray}
{\rm Tr}e^{t{\cal M}_{\rm FP}}
&=&\tr\int d^3x\int{d^3p\over(2\pi)^3}e^{-p^2t}
\left\{
1 + \left(\DD^2 + [\pp]^2\right)t
-4p_ip_j\DD_i\DD_j{t^2\over2!}
\right.
\nonumber \\ &&
\qquad\qquad\qquad
\left.
+\left(\DD^2\left(\DD^2+[\pp]^2\right)
      +[\pp]^2\left(\DD^2+[\pp]^2\right)\right)
{t^2\over2!}
\right.
\nonumber \\ &&
\qquad\qquad\qquad
\left.
-4p_ip_j\left(\left(\DD^2+[\pp]^2\right)\DD_i\DD_j
             +\DD_i\left(\DD^2+[\pp]^2\right)\DD_j
\right.\right.
\nonumber \\ &&
\qquad\qquad\qquad
\left.\left.
	     +\DD_i\DD_j\left(\DD^2+[\pp]^2\right)\right)
{t^3\over 3!}
+16p_ip_jp_kp_l\DD_i\DD_j\DD_k\DD_l
{t^4\over 4!} \right.
\nonumber \\ &&
\qquad\qquad\qquad
\left. + O(\DD^6)
\right\}
\nonumber \\ &&
\nonumber \\ &&
={\pi^{3/2}\over(2\pi)^3}\tr\int d^3x
\left\{
t^{-3/2} + [\pp]^2t^{-1/2}
+{1\over 6}
\left(-\DD_i\DD^2\DD_i+\DD_i\DD_j\DD_i\DD_j
\right.
\right.
\nonumber \\ &&
\qquad\qquad\qquad\qquad
\left.
\left.
+\DD^2[\pp]^2-2\DD_i[\pp]^2\DD_i+[\pp]^2\DD^2
+3[\pp]^4
\right)
t^{1/2}\right.
\nonumber \\ &&
\qquad\qquad\qquad\qquad
\left.+ O(t^{3/2})
\right\}\;,
\label{fadr}
\end{eqnarray}
where, for simplicity, we already supressed all terms with an odd
number of $p_i$'s because they trivially vanish once the
$d^3p$-integration is performed. Furthermore we introduced the
shorthand notation $[X]$ which stands for
\begin{equation}
[X]^{ab}\equiv f^{acb}X^c
\label{no}
\end{equation}
and should not be confused with the commutator notation $[\DD_i,\DD_j]$.

The evaluation of (\ref{fadr}) goes straightforward. One easily verifies
that
\begin{equation}
\tr\left(\DD_i\DD_j\DD_i\DD_j-\DD_i\DD^2\DD_i\right)
={1\over2}\tr\left([\DD_i,\DD_j][\DD_i,\DD_j]\right)
=-{N\over2}\FF^a_{ij}\FF^a_{ij}\;,
\label{si}
\end{equation}
and
\begin{eqnarray}
\lefteqn{\tr\left(\DD^2[\pp][\pp]+[\pp][\pp]\DD^2-2\DD_i[\pp][\pp]\DD_i\right)
=} &&
\nonumber \\ &&
 \tr\left([\DD\DD\pp][\pp]+[\pp][\DD\DD\pp]+2[\DD\pp][\DD\pp]\right)
\nonumber \\ &&
=-2N\left((\DD_i\pp)^a(\DD_i\pp)^a
+ {\lambda\over 2e^2}\left(\pp^{a2}-1\right)\pp^{b2}\right)\!.
\label{sii}
\end{eqnarray}
Substituting these expressions into (\ref{fadr}) and performing the remaining
traces we arrive at
\begin{eqnarray}
{\rm Tr}e^{t{\cal M}_{\rm FP}}
&=&{N\pi^{3/2}\over(2\pi)^3}\int d^3x
\left\{
{N^2-1\over N}t^{-3/2}- \pp^{a2}t^{-1/2}
\right.
\nonumber \\ &&
\qquad\qquad \left.
-{1\over3}
\left(
{1\over4}\FF_{ij}^a\FF_{ij}^a
+(\DD_i\pp)^a(\DD_i\pp)^a
+{\lambda\over 2e^2}\left(\pp^{a2}-1\right)\pp^{b2}
\right.
\right.
\nonumber \\ &&
\qquad\qquad
\left.
\left.
-{3\over N}\left(\pp^{a2}\right)^2
-{3\over4}\left(d^{abc}\pp^b\pp^c\right)^2
\right)t^{1/2}+ O(t^{3/2})
\right\}.
\label{fadd}
\end{eqnarray}
The corresponding expression for the 'free' Faddeev-Popov matrix,
${\cal M}^0_{\rm FP}$ is immediately obtained by setting $\AA_i^a=0$
and $\pp^a=\vv^a=v^a/v$. Thus
\begin{eqnarray}
\lefteqn{{\rm Tr}\left(e^{{\cal M}_{\rm FP}t}-e^{{\cal M}^0_{\rm FP}t}\right)
=}&&
\nonumber \\ &&
={N\pi^{3/2}\over(2\pi)^3}\int d^3x
\left\{
- \left(\pp^{a2}-1\right)t^{-1/2}
-{1\over3}
\left(
{1\over4}\FF_{ij}^a\FF_{ij}^a
+(\DD_i\pp)^a(\DD_i\pp)^a
\right.
\right.
\nonumber \\ &&
\qquad\qquad\qquad\qquad
\left.
\left.
+{\lambda\over 2e^2}\left(\pp^{a2}-1\right)\pp^{b2}
-{3\over N}\left(\left(\pp^{a2}\right)^2-1\right)
\right.
\right.
\nonumber \\ &&
\qquad\qquad\qquad\qquad
\left.
\left.
-{3\over4}\left(\left(d^{abc}\pp^b\pp^c\right)^2
-\left(d^{abc}\vv^b\vv^c\right)^2\right)
\right)t^{1/2}
+ O(t^{3/2})
\right\}\;.
\nonumber \\ &&
\label{faddd}
\end{eqnarray}

Since the operators ${\cal M}_{\rm FP}$ and ${\cal M}^0_{\rm FP}$ are
positive definite and possess a continuous spectrum we can employ
the heat kernel approximation. Thereby we assume that the left-hand
side of (\ref{faddd}) is a rapidly decaying function of $t$ which can be
approximated by the first few terms of its expansion. Thus, to obtain an
estimate for the $t$-integral in (\ref{fadr}) we insert the
expansion (\ref{faddd}) into (\ref{fadr}) and replace the infinite upper
limit of
the integral by a finite limit $t_0$. For $t_0$ we choose that value
at which the result as a function of $t_{0}$ possesses an extremum.
In this way we obtain
\begin{equation}
\ln\left(
{\det[-{\cal M}_{\rm FP}]\over \det[-{\cal M}^0_{\rm FP}]}
\right)_{\rm reg} \approx
-\lim_{s\to0}\left\{{d\over ds}\left({M^{2s}\over\Gamma(s)}
\int_0^{t_0} dt t^{s-1}
\left[\alpha_{{\cal M}_{\rm FP}} t^{-1/2}
-\beta_{{\cal M}_{\rm FP}} t^{1/2}\right]
\right)
\right\}\;,
\label{fadrr1}
\end{equation}
where
\begin{eqnarray}
\alpha_{{\cal M}_{\rm FP}}
& = & {N\pi^{3/2}\over(2\pi)^3}\int d^3x
\left(1-\pp^{a2}\right)\;,
\nonumber \\
\nonumber \\
\beta_{{\cal M}_{\rm FP}} &= &{N\pi^{3/2}\over3(2\pi)^3}\int d^3x
\left(
{1\over4}\FF_{ij}^a\FF_{ij}^a
+(\DD_i\pp)^a(\DD_i\pp)^a
+{\lambda\over 2e^2}\left(\pp^{a2}-1\right)\pp^{b2}
\right.
\nonumber \\ &&
\qquad\qquad\qquad
\left.
-{3\over N}\left(\left(\pp^{a2}\right)^2-1\right)
-{3\over4}\left(\left(d^{abc}\pp^b\pp^c\right)^2
-\left(d^{abc}\vv^b\vv^c\right)^2\right)
\right)
\nonumber \\ &&
\label{coef}
\end{eqnarray}
and $t_0$, determined by the extremum condition
$\alpha_{{\cal M}_{\rm FP}}t_{0}^{-1/2}
- \beta_{{\cal M}_{\rm FP}}t_{0}^{1/2} = 0$ is given by
\begin{equation}
t_{0} ={\alpha_{{\cal M}_{\rm FP}}
\over\beta_{{\cal M}_{\rm FP}}}\;.
\label{ttt}
\end{equation}
Note, that $t_{0}$ must be positive. Otherwise we must take
into account the next higher order in the $t$-expansion. In fact, we can
easily convince ourselves that $t_{0}$ indeed is always positive. Since
$\pp^{a2}\leq 1$ the coefficient $\alpha_{{\cal M}_{\rm FP}}$ is clearly
positive. As far as $\beta_{{\cal M}_{\rm FP}}$ is concerned the only
negative term is the third one which, according to the field equation for
$\pp$ will exactly be canceled by the second term. Consequently, both
coefficients are always positive.

Finally we have to perform the integral over $dt$ and to take the limit
$s\to 0$ which leads to the following result:
\begin{equation}
\ln\left(
{\det[-{\cal M}_{\rm FP}]\over \det[-{\cal M}^0_{\rm FP}]}
\right)_{\rm reg}
\approx 4\sqrt{\alpha_{{\cal M}_{\rm FP}}\beta_{{\cal M}_{\rm FP}}}
= 4\beta_{{\cal M}_{\rm FP}}t_0^{1/2}\;.
\label{fadrr}
\end{equation}

Apart from the fact that the expressions become rather lengthy the
computation of the quantum fluctuation determinant, i.e.,
\begin{equation}
\ln\left(
{\widetilde{\det}[{\cal M}]\over {\det}[{\cal M}^0]}
\right)_{\rm reg} =
-\lim_{s\to0}\left\{{d\over ds}\left({M^{2s}\over\Gamma(s)}
\int_0^\infty dt t^{s-1}
{\rm Tr}\left(e^{t{\cal M}}-e^{t{\cal M}^0}-n_{\cal M}\right)
\right)\right\}\;,
\label{fadm}
\end{equation}
goes exactly the same way. Instead of (\ref{fadr}) one finds
\begin{eqnarray}
\lefteqn{{\rm Tr}e^{{\cal M}t}=} &&
\nonumber \\ &&
=\tr\int d^3x\int{d^3p\over(2\pi)^3}e^{-p^2t}
\left\{
(\delta_{ij}+1)
\right.
\nonumber \\ &&
\qquad\qquad\qquad\left.
+\left((\DD^2 + [\pp]^2)\delta_{ij}+2[\FF_{ij}]
+\DD^2-V(\pp)+[\pp]^2\right)t
\right.
\nonumber \\ &&
\qquad\qquad\qquad
\left.
-4p_kp_l\DD_k\DD_l(\delta_{ij}+1){t^2\over2!}
+\left(
(\DD^2+[\pp]^2)^2\delta_{ij}
+2[\FF_{ij}](\DD^2+[\pp]^2)
\right.
\right.
\nonumber \\ &&
\qquad\qquad\qquad
\left.
\left.
+2(\DD^2+[\pp]^2)[\FF_{ij}]
+4[\FF_{ik}][\FF_{kj}]
-4[\DD_i\pp][\DD_j\pp]-4[\DD_k\pp]^2
\right.
\right.
\nonumber \\ &&
\qquad\qquad\qquad
\left.
\left.
+(\DD^2-V(\pp)+[\pp]^2)^2
\right)
{t^2\over2!}
-4p_kp_l
\left(
      \DD_k\DD_l(\DD^2+[\pp]^2)\delta_{ij}
\right.
\right.
\nonumber \\ &&
\qquad\qquad\qquad
\left.
\left.
      +\DD_k(\DD^2+[\pp]^2)\DD_l\delta_{ij}
      +(\DD^2+[\pp]^2)\DD_k\DD_l\delta_{ij}
      +2\DD_k\DD_l[\FF_{ij}]
\right.
\right.
\nonumber \\ &&
\qquad\qquad\qquad
\left.
\left.
      +2\DD_k[\FF_{ij}]\DD_l
      +2[\FF_{ij}]\DD_k\DD_l
      +\DD_k\DD_l(\DD^2-V(\pp)+[\pp]^2)
\right.
\right.
\nonumber \\ &&
\qquad\qquad\qquad
\left.
\left.
      +\DD_k(\DD^2-V(\pp)+[\pp]^2)\DD_l
      +(\DD^2-V(\pp)+[\pp]^2)\DD_k\DD_l
\right)
{t^3\over 3!}
\right.
\nonumber \\ &&
\qquad\qquad\qquad
\left.
+16p_kp_lp_mp_n\DD_k\DD_l\DD_m\DD_n(\delta_{ij}+1)
{t^4\over 4!} + O(\DD^6)
\right\}
\nonumber \\ &&
\nonumber \\ &&
={\pi^{3/2}\over(2\pi)^3}\tr\int d^3x
\left\{
4t^{-3/2} + \left(4[\pp]^2-V(\pp)\right)t^{-1/2}
\right.
\nonumber \\ &&
\qquad\qquad\qquad
\left.
+{1\over 6}
\left(
4\left(\DD_i\DD_j\DD_i\DD_j-\DD_i\DD^2\DD_i\right)
-12[\FF_{ij}][\FF_{ij}]
-24[\DD_i\pp][\DD_i\pp]
\right.
\right.
\nonumber \\ &&
\qquad\qquad\qquad\qquad\qquad
\left.
\left.
+4\left(\DD^2[\pp]^2-2\DD_i[\pp]^2\DD_i+[\pp]^2\DD^2+3[\pp]^4\right)
\right.
\right.
\nonumber \\ &&
\qquad\qquad\qquad\qquad\qquad
\left.
\left.
-\left(\DD^2V(\pp)-2\DD_iV(\pp)\DD_i+V(\pp)\DD^2-3V(\pp)^2\right)
\right.
\right.
\nonumber \\ &&
\qquad\qquad\qquad\qquad\qquad
\left.
\left.
-3\left([\pp]^2V(\pp)+V(\pp)[\pp]^2\right)
\right)
t^{1/2}
+ O(t^{3/2})
\right\}\;,
\nonumber \\ &&
\label{mat}
\end{eqnarray}
where, in order to simplify the expression at least a little bit, we
used the shorthand notation $V(\pp)$ for
\begin{equation}
V^{ab}(\pp)\equiv {\lambda\over 2e^2}
\left(\pp^{c2}-1\right)\delta^{ab}
+{\lambda\over e^2}\pp^a\pp^b\;.
\label{noo}
\end{equation}
With the help of (\ref{si}), (\ref{sii}) and
\begin{eqnarray}
\lefteqn{\tr\left(\DD^2V(\pp)+V(\pp)\DD^2-2\DD_iV(\pp)\DD_i\right)
=} &&
\nonumber \\ &&
\qquad\qquad\qquad
= {\left(N^2+1\right)\lambda\over e^2}
\left((\DD_i\pp)^a(\DD_i\pp)^a+\pp^a(\DD_i\DD_i\pp)^a\right)
\nonumber \\ &&
\qquad\qquad\qquad
={\left(N^2+1\right)\lambda\over e^2}
\left((\DD_i\pp)^a(\DD_i\pp)^a
      +{\lambda\over 2e^2}\left(\pp^{a2}-1\right)\pp^{b2}\right)
\label{siii}
\end{eqnarray}
we obtain from (\ref{mat})
\begin{eqnarray}
\lefteqn{{\rm Tr}e^{{\cal M}t}=} &&
\nonumber \\ &&
={\pi^{3/2}\over(2\pi)^3}\int d^3x
\left\{
4\left( N^2-1\right)t^{-3/2}
\right.
\nonumber \\ &&
\qquad\qquad\qquad
\left.
-\left(4N\pp^{a2}+{\lambda\over 2e^2}
\left(\pp^{a2}-1\right)\left(N^2-1\right)
       +{\lambda\over e^2}\pp^{a2}\right)t^{-1/2}
\right.
\nonumber \\ &&
\qquad\qquad\qquad
\left.
+{1\over6}
\left(
10N\FF_{ij}^a\FF_{ij}^a
+16N(\DD_i\pp)^a(\DD_i\pp)^a
-{N\lambda\over e^2}\left(\pp^{a2}-1\right)\pp^{b2}
\right.
\right.
\nonumber \\ &&
\qquad\qquad\qquad\quad
\left.
\left.
-(N^2+1){\lambda\over e^2}(\DD_i\pp)^a(\DD_i\pp)^a
+24\left(\pp^{a2}\right)^2
+6N\left(d^{abc}\pp^b\pp^c\right)^2
\right.
\right.
\nonumber \\ &&
\qquad\qquad\qquad\quad
\left.
\left.
+{\lambda^2\over 4e^4}
\left(
\left(N^2-1\right)\left(\pp^{a2}-1\right)\left(\pp^{b2}-3\right)
\right.
\right.
\right.
\nonumber \\ &&
\qquad\qquad\qquad\qquad\qquad\quad
\left.
\left.
\left.
+8\left(\pp^{a2}-1\right)\pp^{b2}+12\left(\pp^{a2}\right)^2
\right)
\right)t^{1/2}
+ O(t^{3/2})
\right\}.
\nonumber \\ &&
\label{matt}
\end{eqnarray}
If we substitute this expression and the corresponding one
for ${\cal M}^0$ into (\ref{fadm}) and remember that $\cal M$ has
$n_{\cal M}=3$ translational zero modes we arrive at
\begin{equation}
\ln\left(
{\widetilde{\det}[{\cal M}]\over {\det}[{\cal M}^0]}
\right)_{\rm reg} \approx
-\lim_{s\to0}\left\{{d\over ds}\left({M^{2s}\over\Gamma(s)}
\int_0^{t_1} dt t^{s-1}
\left[\alpha_{\cal M} t^{-1/2}
+\beta_{\cal M} t^{1/2}-3\right]
\right)
\right\},
\label{fadmm}
\end{equation}
where
\begin{eqnarray}
\alpha_{\cal M}&=&{\pi^{3/2}\over(2\pi)^3}
\left(4N+{\lambda\over 2e^2}\left(N^2+1\right)\right)\int d^3x
\left(1-\pp^{a2}\right)\;,
\nonumber \\
\nonumber \\
\beta_{\cal M}&=&{\pi^{3/2}\over6(2\pi)^3}\int d^3x
\left(
10N\FF_{ij}^a\FF_{ij}^a
+16N(\DD_i\pp)^a(\DD_i\pp)^a
-{N\lambda\over e^2}\left(\pp^{a2}-1\right)\pp^{b2}
\right.
\nonumber \\ &&
\qquad\qquad\qquad\quad
\left.
-(N^2+1){\lambda\over e^2}(\DD_i\pp)^a(\DD_i\pp)^a
+24\left(\left(\pp^{a2}\right)^2-1\right)
\right.
\nonumber \\ &&
\qquad\qquad\qquad\quad
\left.
+{\lambda^2\over 4e^4}\left(\pp^{a2}-1\right)
\left(\left(N^2-1\right)\left(\pp^{b2}-3\right)
+4\left(5\pp^{b2}+3\right)
\right)
\right.
\nonumber \\ &&
\qquad\qquad\qquad\quad
\left.
+6N\left(\left(d^{abc}\pp^b\pp^c\right)^2
-\left(d^{abc}\vv^b\vv^c\right)^2\right)
\right)\;.
\label{coeff}
\end{eqnarray}
The expression for $t_1$ follows from the extremum condition
$\alpha_{\cal M}t_1^{-1/2}+\beta_{\cal M}t_1^{1/2}-3=0$.
Since $\pp^2\leq 1$ the coefficient $\alpha_{\cal M}$
is always positive and therefore
\begin{eqnarray}
t^{1/2}_1=\cases{{3\over2\beta_{\cal M}}+
\sqrt{{9\over4\beta_{\cal M}^2}-{\alpha_{\cal M}\over\beta_{\cal M}}},
&if $\beta_{\cal M} <0$;\cr\cr
{3\over2\beta_{\cal M}}-
\sqrt{{9\over4\beta_{\cal M}^2}-{\alpha_{\cal M}\over\beta_{\cal M}}},
&if $\beta_{\cal M} >0$.\cr}
\label{tmo}
\end{eqnarray}

Since $t^{1/2}_1$ must be real and positive $\alpha_{\cal M}$ and
$\beta_{\cal M}$ must satisfy the inequality
$\alpha_{\cal M}\beta_{\cal M}\leq 9/4$. Otherwise the next higher order
in the $t$-expansion must be taken into account. For $N=2$, i.e., for
the $SU(2)$ gauge model, we have checked explicitly that the inequality
holds. But so far we do not have a general proof that this inequality
holds for monopole configurations of arbitrary $SU(N)$ gauge groups.

Under the assumption that the above inequality holds for arbitrary $N$
we can integrate over $dt$ and take the limit $s\to 0$. This then gives
\begin{equation}
\ln\left(
{\widetilde{\det}[{\cal M}]\over \widetilde{\det}[{\cal M}^0]}
\right)_{\rm reg} \approx
3\gamma_E +2\alpha_{\cal M}t_1^{-1/2}-2\beta_{\cal M} t_1^{1/2}
+ 3\ln\left(M^2t_1\right)\;.
\label{fadmmm}
\end{equation}
where $\gamma_E\approx 0.5772...$ is the Euler constant.

If we insert (\ref{fadrr}) and (\ref{fadmmm}) into (\ref{ZZZZ}) we
obtain the following expression for the one-monopole partition function
\begin{eqnarray}
Z_1&=&\left({m_W\over e^2}\right)^{3/2}
\int d{\bf R}\; M^3
{\cal N}^{3/2}\exp\left(-S_{\rm m}[\AA^a_i,\pp^a]\right)
\nonumber \\ &&
\nonumber \\ &&
\qquad\quad\times
\exp\left[4\beta_{{\cal M}_{\rm FP}}t_0^{1/2}\right]
\exp\left[{3\over2}\gamma_E -\alpha_{\cal M}t_1^{-1/2}
+\beta_{\cal M} t_1^{1/2}-
{3\over2}\ln\left(M^2t_1\right)\right]
\nonumber \\ &&
\nonumber \\ &&
=\left({m_W\over e^2/4\pi}\right)^{3/2}
\int d{\bf R}\; {1\over2\pi} A(\lambda/e^2)
\exp\left[-{m_W\over e^2/4\pi}C(\lambda/e^2)\right]\;,
\label{ZZZZZ}
\end{eqnarray}
where
\begin{equation}
C(\lambda/e^2)
=\left({m_W\over e^2/4\pi}\right)^{-1} S_m[\AA^a_i, \pp^a]
\label{ohjeA}
\end{equation}
is the usual one-monopole mass integral and
\begin{equation}
A(\lambda/e^2)=2\pi\left({{\cal N}\over 4\pi}\right)^{3/2}
{\exp\left[4\beta_{{\cal M}_{\rm FP}}t_0^{1/2}
+\beta_{\cal M}t_1^{1/2}
-\alpha_{\cal M}t_1^{-1/2}
-{3\over2}\gamma_E\right]\over t_1^{3/2}}
\;.
\label{ohjeB}
\end{equation}
Note, that in contrast to the corresponding result for the one-loop
instanton partition function in four dimensions \cite{dyak} the
Pauli-Villars regulator mass $M$ drops out from the final expression.

Having an explicit expression for one-monopole partition function at
hand we can now study the dilute monopole gas. The simplest example
of such a gas is the $SU(2)$ monopole gas consisting of
't Hooft-Polyakov monopoles \cite{poly}.

\vspace{1.0cm}
\section{The $SU(2)$ dilute monopole gas}

For the $SU(2)$ gauge group and $\lambda/e^2>0$ it has been shown
that if the mean distance between two widely separated monopoles
($\vert{\bf R}_i -{\bf R}_j\vert>\!>1$) is large compared to
$(\lambda/e^2)^{-1}$ the superposition principle holds and the
interaction between the monopoles is described by the Coulomb
interaction \cite{magruder}. Thus in this case the action of a system
of $n$ monopoles with magnetic charges $q_i=\pm1$ can be written in
the form
\begin{equation}
S_n = S_{\rm m}\sum_iq_i^2 + {\pi m_W\over2e^2}\sum_{i\not=j}
{q_iq_j\over\vert{\bf R}_i -{\bf R}_j\vert}
\label{sn}
\end{equation}
and the partition function $Z$ for a gas of monopoles becomes
\begin{equation}
Z=\sum_{n}{\left(z_1\right)^n\over n!}
\int \prod_{k=1}^n
d {\bf R}_k
\exp\left(-{\pi m_W\over2e^2}\sum_{i\not=j}
{q_iq_j\over\vert{\bf R}_i -{\bf R}_j\vert}\right)
\label{Zpuren}
\end{equation}
with (cf. (\ref{ZZZZZ}))
\begin{equation}
z_1={1\over2\pi}
\left({m_W\over (e^2/4\pi)}\right)^{3/2} A(\lambda/e^2)
\exp\left[-{m_W\over e^2/4\pi}C(\lambda/e^2)\right]\;.
\label{zzz}
\end{equation}
Note, that in the Prasad-Sommerfield limit $\lambda/e^2\to0$ there appears
a change in the interaction of monopoles \cite{mantona, mantonb, mantonc}.
The Higgs field
becomes massless, and the attractive force associated with it becomes
long range. This force cancels the repulsive magnetic force between like
charges and doubles the attractive force between unlike charges.
Consequently two Prasad-Sommerfield monopoles do not interact via the
Coulomb force if they possess the same magnetic charge.

The Coulomb gas of magnetic monopoles interacting with an external
magnetic charge $\rho({\bf x})$ is described by the partition function
\begin{eqnarray}
Z[\eta]&=&\sum_{n}{\left(z_1\right)^n\over n!}
\int \prod_{k=1}^n
d {\bf R}_k
\exp\left(-{\pi m_W\over2e^2}\sum_{i\not=j}
{q_iq_j\over\vert{\bf R}_i -{\bf R}_j\vert}\right)
\nonumber \\ &&
\nonumber \\ &&
\qquad\qquad\qquad\times
\exp\left(i\int d^3x\sum_{i}q_i\delta({\bf x}
-{\bf R_i})\eta({\bf x})\right)
\label{Zn}
\end{eqnarray}
where $\eta$ is the potential corresponding to the external magnetic
charge, i.e., $\Delta\eta({\bf x})=2\pi\rho({\bf x})$. A path integral
expression for $Z[\eta]$ is now easily obtained by rewriting the Coulomb
interaction in (\ref{Zn}) as an integral over an auxiliary field
$\chi({\bf R})$. The resulting expression reads
\begin{equation}
Z[\eta] = \int{\cal D}[\chi]\exp\left[-{e^2\over4\pi^2m_W}
\int d^3x\left({1\over2}(\nabla(\chi-\eta))^2
- \omega^2\cos\chi\right)\right]
\label{zn}
\end{equation}
where $\omega$ given by
\begin{equation}
\omega^2 = 8\pi^2{m_W\over e^2}z_1
=\left({m_W\over (e^2/4\pi)}\right)^{5/2}A(\lambda/e^2)
\exp\left[-{m_W\over e^2/4\pi}C(\lambda/e^2)\right]
\label{znn}
\end{equation}
is the gauge boson mass in the unbroken $U(1)$ gauge sector in units
of $m_W$, that is, $m_{U(1)}=\omega m_W$.

The Euclidean functional integral (\ref{zn}) describes a Debye plasma
of monopoles and antimonopoles if the Debye radius given by $1/m_{U(1)}$
is large compared to the mean distance between the monopoles. If $n$ is
the density of monopoles the mean distance is proportional to $n^{-1/3}$.
The density of monopoles on the other hand is proportional to
$(e\,m_{U(1)})^2$. Thus the condition for the Debye approximation
is
\begin{equation}
{m_{U(1)}\over e^2} <\!<1\;.
\label{debye}
\end{equation}
Since $m_{U(1)}$ is proportional to
$\exp[-m_W C(\lambda/e^2)/2(e^2/4\pi)]$ it is clear that by taking
$(e^2/4\pi)$ small the monopole gas can be made as dilute as one
likes. The nonlinearities in functional integral (\ref{zn})
become exponentially small, too, because $m_{U(1)}/e^2$ is nothing
but the dimensionless effective coupling of the monopole gas. However,
because of the long-range magnetic Coulomb interaction between
monopoles, the gas does not become noninteracting even for very
small densities. This is a general feature of the three
dimensional compact $U(1)$ gauge model.

The ''area law'' behavior follows from the partition function $Z[\eta]$
if one  considers an external monopole charge density $\rho_l({\bf x})$
generated by an Wilson loop $l$ in the $x_1$, $x_2$ plane. For a large
Wilson loop $l$ $\rho_l({\bf x})$ is approximately given by the monopole
charge density which is generated by an classical external electric
current loop
$j_i=\epsilon_{ijk}\partial_jH_k$ with $\partial_jH_j=\rho_l$, i.e.,
\begin{equation}
\rho_l({\bf x})=\delta'(x_3)\Theta_l(x_1,x_2)\;,
\label{density}
\end{equation}
where $\Theta(x_1,x_2)=1$ if $x_1$ and $x_2$ are the coordinates of a
point inside the loop $l$ and zero otherwise.

In the semi-classical approximation $Z[\eta_l]$ is approximated by
\begin{equation}
Z[\eta_l] \approx \int{\cal D}[\chi]
\exp\left[-{e^2\over4\pi^2m_W}
\int d^3x\left({1\over2}(\nabla(\chi_{\rm cl}-\eta_l))^2
- \omega^2\cos\chi_{\rm cl}\right)\right]\;,
\label{zncl}
\end{equation}
where $\chi_{\rm cl}$ is a solution of
\begin{equation}
\nabla^2\chi_{\rm cl}=\nabla^2\eta_l+\omega^2\sin\chi_{\rm cl}
=2\pi\rho_l+\omega^2\sin\chi_{\rm cl}\;.
\label{clss}
\end{equation}
An approximate solution of (\ref{clss}) with $\rho_l$ given
by (\ref{density}) is
\begin{eqnarray}
\chi_{\rm cl}=\cases{
4\Theta_l(x_1,x_2)\arctan\left(e^{-\omega x_3}\right),& if $x_3>0$,\cr\cr
4\Theta_l(x_1,x_2)\arctan\left(e^{\omega x_3}\right),& if $x_3<0$.\cr}
\label{solclas}
\end{eqnarray}
If we insert this solution into (\ref{zncl}) we finally arrive at
\begin{equation}
Z[\eta_l]\approx e^{-\sigma S_l}
\label{zstr}
\end{equation}
where $S_l$ is the area enclosed by the loop $l$ and
\begin{equation}
\sigma={2e^2\over\pi^2}m_W\omega={2e^2\over\pi^2}m_{U(1)}
\label{string}
\end{equation}
the string tension. This is Polyakov's formula (corrected for misprints).

To summarize, the polarization of a dilute monopole gas of 't Hooft-Polyakov
monopoles screens the long range magnetic Coulomb force so that an infrared
stable semiclassical expansion becomes possible. There are no massless
particles in the theory and non-singlet states are confined by a linearly
rising potential. The $U(1)$ gauge boson mass and the string tension of the
dilute gas are given by
\begin{equation}
{m_{U(1)}\over m_W}
=\omega\left({m_W\over e^2/4\pi)},{\lambda\over e^2}\right)\;,
\label{ratioa}
\end{equation}
and
\begin{equation}
{\sigma\over m_W^2}=8\pi\left({m_W\over e^2/4\pi)}\right)^{-1}
\omega\left({m_W\over e^2/4\pi)},{\lambda\over e^2}\right)\;,
\label{ratiob}
\end{equation}
where
\begin{equation}
\omega^2\left({m_W\over e^2/4\pi)},{\lambda\over e^2}\right)
=\left({m_W\over (e^2/4\pi)}\right)^{5/2}A(\lambda/e^2)
\exp\left[-{m_W\over e^2/4\pi}C(\lambda/e^2)\right].
\label{znnsum}
\end{equation}

\vspace{1.0cm}
\section{Numerical results for the SU(2) gauge model}

The monopole configuration of the $SU(2)$ gauge model is of the form
\cite{mona}
\begin{eqnarray}
\vec{\AA}^a(\vec r)\sigma^a &=& \vec\sigma\times\vec r\,
{h(r)\over r}\;,
\nonumber \\
\nonumber \\
\pp^a(\vec r)\sigma^a &=& \vec r\cdot\vec\sigma\,
{g(r)\over r}\;,
\label{fields}
\end{eqnarray}
where $\sigma^a$ are the Pauli matrices. The functions $h(r)$ and $g(r)$
are obtained by minimizing the action (\ref{ss}), that is the monopole mass
integral
\begin{eqnarray}
C(\lambda/e^2)&=&\int_0^\infty\!dr\left(
\left(rh'(r)+h(r)\right)^2
+2h^2(r)\left(1+{rh(r)\over2}\right)^2\right.
\nonumber \\
\nonumber \\ &&
\qquad\qquad\left.
+{r^2\over2}g'^2(r)+g^2(r)\left(1+rh(r)\right)^2
+{\lambda\over 8e^2}r^2\left(g^2(r)-1\right)^2\right)\;.
\nonumber \\ &&
\label{ff}
\end{eqnarray}
The corresponding differential equations are
\begin{eqnarray}
&&r^2h''(r)+2rh'(r)-2h(r)-rg^2(r)\left(1+rh(r)\right)-3rh^2(r)-r^2h^3(r)=0\;,
\nonumber \\
\nonumber \\ &&
r^2g''(r)+2rg'(r)-2g(r)+{\lambda\over2e^2}r^2g(r)-4rh(r)g(r)
\nonumber \\ &&
\qquad\qquad\qquad\qquad\qquad
-2r^2h^2(r)g(r)
-{\lambda\over2e^2}r^2g^3(r)=0
\;,
\nonumber \\
\label{diffa}
\end{eqnarray}
which apart from the Prasad-Sommerfield limit ($\lambda/ e^2 \to 0$)
\cite{pra} can only be solved numerically.

To determine the functions $h(r)$ and $g(r)$ we use instead of (\ref{diffa})
the corresponding system of coupled nonlinear integral equations derived by
Bais and Primack \cite{baispr} which can be solved numerically by simple
iteration. In terms of $h(r)$ and $f(r)=1-g(r)$ these integral equations are
given by
\begin{eqnarray}
h(r)&=&h_0(r)+\int_0^\infty G_h(r,r')\left(A_h(r')+B[h_0(r')]\right)dr'\;,
\nonumber \\
\nonumber \\
f(r)&=&g_0(r)+\int_0^\infty G_f(r,r')\left(A_f(r')+B[f_0(r')]\right)dr'\;,
\label{inteq}
\end{eqnarray}
where
\begin{eqnarray}
G_h(r,r')&=&
\cases{\left({\cosh(r)\over r}-{\sinh(r)\over r^2}\right)
{e^{-r'}(1+1/ r')\over r'},
&if $r'>r$,\cr\cr
\left({\cosh(r')\over r'}-{\sinh(r')\over {r'}^2}\right)
{e^{-r}(1+1/ r)\over r},
&if $r'<r$,\cr}
\nonumber \\
\nonumber \\
G_f(r,r')&=&
\cases{\left({\cosh(\sqrt{\lambda/e^2}r)\over \sqrt{\lambda/e^2}r}-
{\sinh(\sqrt{\lambda/e^2}r)\over \left(\sqrt{\lambda/e^2}r\right)^2}\right)
{e^{-\sqrt{\lambda/e^2}r'}\left(1+1/\left(\sqrt{\lambda/e^2}r'\right)\right)
\over \sqrt{\lambda/e^2}r'},
&if $r'>r$,\cr\cr\cr
\left({\cosh(\sqrt{\lambda/e^2}r')\over \sqrt{\lambda/e^2}r'}-
{\sinh(\sqrt{\lambda/e^2}r')\over \left(\sqrt{\lambda/e^2}r'\right)^2}\right)
{e^{-\sqrt{\lambda/e^2}r}\left(1+1/\left(\sqrt{\lambda/e^2}r\right)\right)
\over \sqrt{\lambda/e^2}r},
&if $r'<r$,\cr}
\nonumber \\
\label{green}
\end{eqnarray}
are the Green's functions of the homogeneous equations corresponding to
(\ref{diffa}),
\begin{eqnarray}
A_h(r)&=&-r(1+r h(r))(f^2(r)-2 f(r))-3r h^2(r)-r^2h^3(r)-r,
\nonumber \\
\nonumber \\
A_f(r)&=&2+2r h(r)(2+r h(r))(1-f(r))+{\lambda\over 2e^2}r^2 f^2(r)(3-f(r)),
\label{funka}
\end{eqnarray}
are the nonlinear sources and
\begin{eqnarray}
B[h_0(r)]&=&r^2h_0''(r)+2rh_0'(r)-(2+r^2)h_0(r),
\nonumber \\
\nonumber \\
B[f_0(r)]&=&r^2f_0''(r)+2rf_0'(r)-\left(2+{\lambda\over e^2}r^2\right)f_0(r).
\label{funkb}
\end{eqnarray}
The functions $h_0(r)$ and $f_0(r)$ must satisfy the boundary conditions
\begin{eqnarray}
&&f(r)\mathop{\longrightarrow}_{ r\to0}1-c_f(\lambda/e^2)r+O(r^3),
\qquad
h(r)\mathop{\longrightarrow}_{r\to0}c_h(\lambda/e^2)r +O(r^3),
\nonumber \\ &&
\nonumber \\ &&
f(r)\mathop{\longrightarrow}_{r\to\infty}0,\phantom{-c_f(\lambda/e^2)r+O(r^3)}
\qquad
h(r)\mathop{\longrightarrow}_{r\to\infty}-1/r,
\label{boundary}
\end{eqnarray}
but can otherwise arbitrarily be chosen. As in \cite{baispr} we have used
\begin{equation}
h_0(r)=-{r\over r^2+b(\lambda/e^2)}\qquad\hbox{and}\qquad
f_0(r)=1-{r\over(r^2+a(\lambda/e^2))^{1/2}},
\label{startf}
\end{equation}
where $a(\lambda/e^2)$ and $b(\lambda/e^2)$ are to be determined by
minimizing the monopole mass integral $C(\lambda/e^2)$. The functionals
$B[h_0(r)]$ and $B[f_0(r)]$ then take the form
\begin{eqnarray}
B[h_0(r)]&=&{2b(\lambda/e^2)r(3r^2-b(\lambda/e^2))\over (r^2+b(\lambda/e^2))^3}
-(2+r^2)h_0(r)
\nonumber \\
\nonumber \\
B[f_0(r)]&=&{a(\lambda/e^2)r(r^2-2a)\over (r^2+a(\lambda/e^2))^{5/2}}
-\left({\lambda\over e^2}r^2+2\right)f_0(r).
\label{funkbr}
\end{eqnarray}
For a detailed derivation of the coupled set of integral
equations (\ref{inteq}) we refer the reader to ref. \cite{baispr}.

To study the $\lambda/e^2$-dependence of the $U(1)$ gauge boson mass and
the string tension $\sigma$ we have calculated iterative solutions of
the system (\ref{inteq}) for various values of $\lambda/e^2$. The results
for the mass integral $C(\lambda/e^2)$ and $A(\lambda/e^2)$ which together
determine the total $\lambda/e^2$-dependence of $m_{U(1)}$ and $\sigma$ are
plotted in Fig. 1 and  Fig. 2. In addition we have listed a few illustrative
values in Table 1.

\vglue 1.0truecm
\centerline{
\vbox{\tabskip=0pt \offinterlineskip
\def\tablerule{\noalign{\hrule}}
\halign to 420pt{\strut#&\vrule#\tabskip=1em plus 2em&
\hfil#& \vrule#& \hfil#\hfil& \vrule#&
\hfil#& \vrule#&
%\hfil#& \vrule#&
\hfil#& \vrule#&
\hfil#& \vrule#&
\hfil#& \vrule#\tabskip=0pt\cr\tablerule
\omit&height2pt&
\omit&height2pt&
\omit&height2pt&
\omit&height2pt&
%\omit&height2pt&
\omit&height2pt&
\omit&height2pt&
\omit&height2pt\cr
&&\omit\hidewidth $\lambda/e^2$\hidewidth&&
  \omit\hidewidth $  0.1$\hidewidth&&
  \omit\hidewidth $  0.5$\hidewidth&&
%  \omit\hidewidth $ 1.0$\hidewidth&&
  \omit\hidewidth $  2.0$\hidewidth&&
  \omit\hidewidth $ 10.0$\hidewidth&&
  \omit\hidewidth $100.0$\hidewidth&\cr
\omit&height2pt&
\omit&height2pt&
\omit&height2pt&
\omit&height2pt&
%\omit&height2pt&
\omit&height2pt&
\omit&height2pt&
\omit&height2pt\cr
\tablerule
\omit&height2pt&
\omit&height2pt&
\omit&height2pt&
\omit&height2pt&
%\omit&height2pt&
\omit&height2pt&
\omit&height2pt&
\omit&height2pt\cr
&&$C(\lambda/e^2)$&&
$1.106$&&
$1.119$\phantom{xx}&&
%$1.237$&&
$1.291$&&
$1.433$\phantom{xx}&&
$1.617$\phantom{xx}&\cr
\omit&height2pt&
\omit&height2pt&
\omit&height2pt&
\omit&height2pt&
%\omit&height2pt&
\omit&height2pt&
\omit&height2pt&
\omit&height2pt\cr
\tablerule
\omit&height2pt&
\omit&height2pt&
\omit&height2pt&
\omit&height2pt&
%\omit&height2pt&
\omit&height2pt&
\omit&height2pt&
\omit&height2pt\cr
&&$A(\lambda/e^2)$&&
$3.7\cdot10^{-19}$&&
$9.2\cdot10^{-3}$&&
%$1.068$&&
$6.576$&&
$1.74\cdot10^{-1}$&&
$1.14\cdot10^{-10} $&\cr
%\omit&height2pt&\omit&height2pt&\omit&height2pt&\omit&height2pt&
%\omit&height2pt&\omit&height2pt&\omit&height2pt&\omit&height2pt\cr
%\tablerule
%\omit&height2pt&\omit&height2pt&\omit&height2pt&\omit&height2pt&
%\omit&height2pt&\omit&height2pt&\omit&height2pt&\omit&height2pt\cr
%&&$10.0$&&$ $&&$ $&&$ $&&$ $&&$ $&&$ $&\cr
\omit&height2pt&
\omit&height2pt&
\omit&height2pt&
\omit&height2pt&
%\omit&height2pt&
\omit&height2pt&
\omit&height2pt&
\omit&height2pt\cr
\tablerule}}
}
\vglue 0.3truecm
\centerline{Table I. $C(\lambda/e^2)$ and $A(\lambda/e^2)$ for
different values of $\lambda/e^2$}
\vglue 1.0truecm

In contrast to the mass integral $C(\lambda/e^2)$ which is a
rather slowly increasing function of $\lambda/e^2$ between
$1.0$ ($\lambda/e^2=0$) and $1.787$ ($\lambda/e^2\to\infty$)
the function $A(\lambda/e^2)$ varies substantially. This strong
$\lambda/e^2$ dependence can formally be understood if one compares
the various terms which contribute to mass integral $C(\lambda/e^2)$
and the coefficients $\alpha_{M_{\rm FP}}$, $\beta_{M_{\rm FP}}$,
$\alpha_{M}$ and $\beta_M$ in the heat kernel expansion.
Expressions that do appear in these coefficients but not in the
mass integral are for example $\int d^3x(1-\pp^2)$ and
$\int d^3x((\pp^2)^2-1)$. In contrast to the similar term
$(\lambda/e^2)\int d^3x(\pp^2-1)^2$ which appears in the mass
integral the size of these expressions depends very much
on the shape of Higgs field which is determined by the ratio
$\lambda/e^2$.

At $\lambda/e^2\approx2.2$ the function $A(\lambda/e^2)$ has a
maximum. Since on the other hand $\lambda/e^2=m^2_{\rm Higgs}/m^2_W$
this result is equivalent with the statement that $A(\lambda/e^2)$
is largest if the square of the Higgs mass is roughly twice as large
as the square of the heavy vector boson mass. For $\lambda/e^2>2.2$
$A(\lambda/e^2)$ decreases again and seems to vanish in the limit
$\lambda/e^2\to\infty$.

The behavior of $A(\lambda/e^2)$ for small and large values of
$\lambda/e^2$ becomes more transparent in Fig. 3 where the logarithm
of $A(\lambda/e^2)$ is shown. For curiosity we have also tried to
fit the curve for small and large values of $\lambda/e^2$ and
arrived at the following result. For small $\lambda/e^2$ the
function $A(\lambda/e^2)$ seems to starts out as
\begin{equation}
A(\lambda/e^2)
\approx 125.66(\lambda/e^2)^{1/3}\exp\left(-{4.65\over \lambda/e^2}\right)\;,
\label{solexps}
\end{equation}
and the behavior for larger $\lambda/e^2$ may be described by
\begin{equation}
A(\lambda/e^2)\approx 100.56{
\exp\left(-1.21(\lambda/e^2)^{2/3}\right)\over(\lambda/e^2)^{1/3}}\;.
\label{solexpl}
\end{equation}
However, at the present stage we should not pay too much attention
to the precise values of the constants in (\ref{solexps}) and
(\ref{solexpl}) since we do not know yet how accurate our approximate
calculation of the determinants really is.

Since $m_{U(1)}$ and $\sigma$ are proportional to
$\sqrt{A(\lambda/e^2)}$ the above results show that both
quantities vanish in the limits $\lambda/e^2\to0$ and
$\lambda/e^2\to\infty$. However, physics is very different
in these limits. Let us consider the case $\lambda/e^2\to\infty$
first. For large values of $\lambda/e^2$ the attractive force
associated with the Higgs field is clearly short range since the
potential term in the classical action (\ref{ss}) forces the Higgs
field to unity, i.e., the Higgs field rises from $0$ at $r=0$ to
$1$ within a distance of the order of $(\lambda/e^2)^{-1}$.
Therefore the interaction between the monopoles should indeed be
very well described by the magnetic Coulomb force. The result
that $m_{U(1)}$ and $\sigma$ become small as we take $\lambda/e^2$
large becomes plausible if we consider the energy of an assembly
of monopoles \cite{magruder}. Classically, this energy is given by
the minimum of the three dimensional action (\ref{act})
\begin{eqnarray}
S[\AA,\pp]
&=&\int d^3x\left({1\over 4}\FF^a_{ij}\FF^a_{ij}
+{1\over 2}(\partial_i \vert\pp\vert)(\partial_i\vert\pp\vert)\right.
\nonumber \\ &&
\qquad\qquad\left.
+{1\over 2}\vert\pp\vert^2(\DD_i \widetilde\Phi)^a(\DD_i \widetilde\Phi)^a
+{\lambda \over 8}\left(\vert\pp\vert^2 - v^2\right)^2\right)\;.
\label{ssrr}
\end{eqnarray}
subject to the constraint that the normalized Higgs field
$\widetilde\Phi^a=\vert\pp\vert^{-1}\pp^a$ satisfies the required
asymptotic properties of magnetic flux and nontrivial homotopy, i.e.,
\cite{Arafune}
\begin{equation}
{1\over2}\epsilon_{ijk}f^{abc}\partial_i\pp^a\partial_j\pp^b\partial_k\pp^c
=4\pi\sum_n q_n\delta({\bf R}-{\bf R}_n)\;.
\label{ara}
\end{equation}
Since according to (\ref{ara}) $\partial_i\widetilde\phi$ possesses a
$1/r$ singularity at the position of each monopole the quantity
$\min_x[(\DD_i \widetilde\Phi)^a(\DD_i \widetilde\Phi)^a]$ should
increase as the monopole density increases. However, it is easy to
see that if we take $\lambda$ large the potential term forces
$\vert\pp\vert^2\approx v^2$ and it will be energetically favorable
for $\min_x[(\DD_i \widetilde\Phi)^a(\DD_i \widetilde\Phi)^a]$
to be small. Thus for large values of $\lambda$ the density of
monopoles is small.

If we go to very small values of $\lambda/e^2$ the situation is
more complicated. First we must check that the superposition
principle is still accurate, i.e., the attractive force associated
with the Higgs field can be neglected. This is the case if the
distance between the monopoles in the gas is large compared to
$(\lambda/e^2)^{-1}$. Since the mean distance between the monopoles
is proportional to $(e\,m_{U(1)})^{-2/3}$ (cf. previous section)
and $m_{U(1)}$ according to the above results is proportional to
$(\lambda/e^2)^{1/3}\exp[-{\rm const.}/(\lambda/e^2)]$ the superposition
principle indeed seems justified even for very small values
of $\lambda/e^2$. To understand the drastic decrease of
$m_{U(1)}$ and $\sigma$ for small values of $\lambda/e^2$ let us
again consider the classical energy of an assembly of monopoles
(\ref{ssrr}) subject to the constraint (\ref{ara}). We assume the
presence of an extremely small monopole density and start to
increase it. As the monopole density increases the quantity
$\min_x[(\DD_i \widetilde\Phi)^a(\DD_i \widetilde\Phi)^a]$
increases. But when this term becomes larger than $\lambda\,v^2$ it
will be energetically favorable for $\vert\pp\vert$ to vanish which
now is possible since the potential term in (\ref{ssrr}) does not
constrain $\vert\pp\vert$ for small values of $\lambda$. But once
$\vert\pp\vert$ vanishes it will be energetically favorable for
the gauge fields to vanish, too.

Indeed a detailed study of the $\lambda/e^2$ dependence of the expressions
$\alpha_{{\cal M}_{\rm FP}}$, $\beta_{{\cal M}_{\rm FP}}$,
$\alpha_{{\cal M}}$ and $\beta_{{\cal M}}$ shows that in the limit
$\lambda/e^2\to\infty$ those terms with the highest power in the
Higgs self-coupling $\lambda/e^2$ become dominant. These are just the
terms which result from the Higgs potential. Since they appear with a
negative sign in the exponential function (\ref{ohjeB})
$A(\lambda/e^2)$ becomes small as $\lambda/e^2$ becomes very large. In
the limit $\lambda/e^2\to0$, on the other hand, it is not so that
$A(\lambda/e^2)$ vanishes because the Higgs coupling becomes small
but because the core of size $1/(\lambda/e^2)$ outside of which the
Higgs field approaches its asymptotic form becomes large.

Thus based on the assumption that for small values of the coupling
$e^2/4\pi$ the dilute monopole gas partition function is a reasonable
approximation to the path integral of the Yang-Mills-Higgs theory the
quantum theory undergoes a smooth transition to a non-confining Higgs
phase with a massless $U(1)$ gauge boson in the limit
$\lambda/e^2\to\infty$ and a rapid but still smooth transition
to a non-confining symmetrical phase in the limit $\lambda/e^2\to0$.
However, for any finite value of $\lambda/e^2$ the quantum theory is
confining and all fields massive.

To illustrate both dependencies namely, that on $e^2/4\pi$ and
$\lambda/e^2$, the $U(1)$ gauge boson mass is shown in Fig. 4a
and Fig. 4b as a function of $e^2/4\pi$ for various values of $\lambda/e^2$.
(The corresponding graphs for the string tension look very similar.)
Since $m_{U(1)}$ is proportional to $\exp[-m_W C(\lambda/e^2)/2(e^2/4\pi)]$
we see a rapid decrease for very small values of $e^2/4\pi$. But we
also see that the curves intersect. The reason for this is that the
$\lambda/e^2$-dependence of $m_{U(1)}$ enters in two different ways,
namely on the one hand via $A(\lambda/e^2)$ and on the other hand via
$\exp[-m_W C(\lambda/e^2)/(e^2/4\pi)]$. Although
the $\lambda/e^2$ dependence of the mass integral is rather
weak as compared to $A(\lambda/e^2)$ it can play a certain role for
sufficiently small values of $e^2/4\pi$. To see this let us
compare $m_{U(1)}$ for two different values of $\lambda/e^2$,
say $c_1$ and $c_2$ with $c_1<c_2<2.2$. For $\lambda/e^2<2.2$
$A(\lambda/e^2)$ is an increasing function of $\lambda/e^2$ whereas
$\exp(-m_W C(\lambda/e^2)/(e^2/4\pi)$ is always a
decreasing function. Thus if $e^2/4\pi$ becomes sufficiently small
$m_{U(1)}(c_1)>m_{U(1)}(c_2)$ but otherwise $m_{U(1)}(c_1)<m_{U(1)}(c_2)$.

\vspace{1.0cm}
\section{Summary}

The numerical results presented in the previous section show that
it is indeed worth studying the dilute monopole gas expansion of
the Yang-Mills-Higgs model in more detail, especially, if one views
the expansion as an approximation of the Euclidean path integral of
this model.

The dilute monopole gas expansion of the Yang-Mills-Higgs model
depends on two parameters, $e^2/4\pi$ and $\lambda/e^2$. The
$e^2/4\pi$-dependence becomes already transparent in the formal
expression for the monopole partition function whereas the
$\lambda/e^2$-dependence can only be seen when both the classical
one-monopole mass integral and the corresponding functional
determinants have been computed explicitly.

In this paper the determinants have been evaluated with the
help of the heat kernel expansion, an approximation scheme
which has been proven successful in a corresponding
calculation for instantons. The result of this calculation is
an analytical expression which we denoted by $A(\lambda/e^2)$
and which is valid for arbitrary $SU(N)$ gauge groups.
However, since apart from the Prasad-Sommerfield limit
($\lambda/e^2\to0$) the monopole solutions are not known
analytically the further evaluation of $A(\lambda/e^2)$ has to
be done numerically.

As an example we have considered the $SU(2)$ gauge group.
Although the mass integral $C(\lambda/e^2)$ is only a rather
weakly increasing function between $1.0$ and $1.787$ its
$\lambda/e^2$ dependence plays a certain role for small
values of the coupling $e^2/4\pi$. The much stronger
$\lambda/e^2$ dependent quantity is $A(\lambda/e^2)$ which
decreases for small and large values of $\lambda/e^2$ and has
a maximum around $\lambda/e^2\approx2.2$. For any finite
value of $\lambda/e^2$ the dilute monopole gas is confining.
However, the string tension and the generated $U(1)$
gauge boson mass are in general very small. In both
limits, $\lambda/e^2\to\infty$ and $\lambda/e^2\to0$, the
system becomes non-confining but for physically different
reasons. For $\lambda/e^2\to\infty$ we see a smooth transition
from the confining phase to the non-confining Higgs phase and
for $\lambda/e^2\to0$ a rapid but still smooth transition from
the confining phase to the non-confining symmetrical phase.

Although we have seen that the superposition principle seems accurate
even for very small values of $\lambda/e^2$ one
would prefer a direct check. However, this would require to
take into account the attractive force associated with the
Higgs field which is by far nontrivial. What can certainly be done
is to examine the accuracy of the heat kernel approximation
used in this paper. To this extent one must compute the next
order, i.e., the fourth order in the $t$-expansion and see whether
these higher corrections alter the present results significantly.

\vspace{1.0cm}
\section*{Acknowledgments}
The author thanks Prof.~J.~Polonyi for helpful discussions and Prof.~E.~Werner
for his continuous encouragement and support.

\vspace{1.0cm}
\section*{Figure Captions}

\begin{description}
\item[Figure 1:] The monopole mass integral as a function of $\lambda/e^2$
\item[Figure 2:] The $\lambda/e^2$-dependence which results from the
computation of the one-loop corrections to the monopole partition function.
\item[Figure 3:] The logarithm of $A(\lambda/e^2)$. A fit for small and
large values of $\lambda/e^2$ leads to an asymptotic behavior as given
in (\ref{solexps}) and (\ref{solexpl}).
\item[Figure 4a:] The logarithm of the $U(1)$ gauge boson mass in units of
$m_w$ as a function of $e^2/4\pi$ in units of $m_W$ for various values of
$\lambda/e^2$;\\ {\vrule width 0.2cm height 0.12truecm depth -0.06truecm}
\hglue 0.02truecm{\vrule width 0.22truecm height 0.12truecm depth -0.06truecm}
\hglue 0.02truecm{\vrule width 0.22truecm height 0.12truecm depth -0.06truecm}
\hglue 0.02truecm{\vrule width 0.22truecm height 0.12truecm depth -0.06truecm}
: $\lambda/e^2=0.1$, $\;\;$
{\vrule width 0.22cm height 0.12truecm depth -0.06truecm}
\hglue 0.02truecm{\vrule width 0.1truecm height 0.12truecm depth -0.06truecm}
\hglue 0.02truecm{\vrule width 0.22truecm height 0.12truecm depth -0.06truecm}
\hglue 0.02truecm{\vrule width 0.1truecm height 0.12truecm depth -0.06truecm}
: $\lambda/e^2=0.4$, $\;\;$
\vrule width 1.5cm height 0.15truecm depth -0.06truecm
: $\lambda/e^2=2$,\\
\vrule width 1.5cm height 0.12truecm depth -0.06truecm
: $\lambda/e^2=15$, $\;\;$
{\vrule width 0.1cm height 0.12truecm depth -0.06truecm}
\hglue 0.02truecm
{\vrule width 0.1truecm height 0.12truecm depth -0.06truecm}
\hglue 0.02truecm{\vrule width 0.1truecm height 0.12truecm depth -0.06truecm}
\hglue 0.02truecm{\vrule width 0.1truecm height 0.12truecm depth -0.06truecm}
\hglue 0.02truecm{\vrule width 0.1truecm height 0.12truecm depth -0.06truecm}
: $\lambda/e^2=100$.
\item[Figure 4b:] See Fig. 4a.
\end{description}

\end{document}